\documentclass[aps,twocolumn,groupedaddress]{revtex4}

\usepackage{graphics,epsfig}

\begin{document}

\title{The anomalous chromomagnetic dipole moment of the top quark in the standard model and beyond}
  
\author{R. Martinez}
\email[romart@ciencias.unal.edu.co]{}
\author{J-Alexis Rodriguez}
\email[alexro@ciencias.unal.edu.co]{}




\affiliation{Departamento de F\'{\i}sica, Universidad Nacional de Colombia\\
Bogota, Colombia}

\begin{abstract}
The anomalous chromomagnetic moment of the top quark arises from one loop
corrections to the vertex $\bar t t g$. We give explicit
formulae for this anomalous coupling in the framework of the standard model,
the two Higgs doublet model and the minimal supersymmetric standard model.
We compare the results for the anomalous coupling with the bound $-0.03 \leq \Delta
\kappa \leq 0.01$ emerging 
from the analysis of $b \to s \gamma$ process with an on-shell
bremsstrahlung gluon. This enables us to
study the allowed region of parameters of the models under consideration.
\end{abstract}
		
\maketitle

The top quark is the heaviest fermion in the standard model (SM) with a mass of
$174 \pm 5.2 $ GeV. Soon  
detailed experimental studies of the top
quark properties, will be available. This will give us
the envisaged high precision of data opening a rich field of new
top quark phenomenology.  In the framework of the SM, the couplings of the
top quark are fixed by the gauge symmetry, the structure of generations  and
the dimension of the interaction Lagrangian. 
This makes the top quark an ideal candidate
to search for new physics  beyond the SM. Anomalous couplings  between
the 
top quark and gauge bosons might affect the 
top quark production at high energies and also its decay rate. 
Precisely measured quantities with virtual top quark
contributions \cite{cheung,nacht,rizzo} will yield further information 
regarding these couplings.

Modifications  of the SM couplings can be traced back to the dimension
of the  
operators in the effective Lagrangian description valid above the electroweak
symmetry breaking scale. They would be, in principle, 
of the same order as the other dimension 5 and 6 couplings 
below the electroweak scale. By means of the dimension 5 coupling to 
an on-shell gluon, the anomalous
chromomagnetic dipole moment of the top quark is defined as
\begin{equation}
L_5=i (\frac {\Delta \kappa}{2}) \frac{g_s}{2 m_t} \bar{u}(t) 
\sigma_{\mu \nu} q^\nu T^a u(t) G^{\mu ,a}
\end{equation}
where $g_s$ and $T^a$ are the $SU(3)_c$ coupling and generators,
respectively, and the parameter $\Delta
\kappa$ is identified with the anomalous chromomagnetic dipole moment of
the top quark. 

The effects due to $\Delta \kappa \neq 0$ were  examined in
flavor physics  
as well as high top quark cross section measurements
\cite{cheung,nacht,rizzo,willen,hikasa}. In the latter 
case,  
the parton level 
differential cross sections of $gg \to \bar t t$  
and 
$\bar q q \to \bar t t$ (the dominant channel at Tevatron
energies) were calculated \cite{rizzo,willen,silver}. The combined
effects of the chromomagnetic and the chromoelectric dipole moment of the top
quark  on the reaction $p \bar p
\to t \bar t X$ were investigated in reference \cite{nacht}. Moreover, previous
analysis has revealed that the differential cross section is sensitive to the
sign of the anomalous chromomagnetic dipole moment 
on account of the interference
with the SM coupling.This can lead to a significant suppression or
enhancement in the production rate \cite{rizzo}. 

 We mention that on the one hand the $\bar{t} t g$ coupling correction to the
total top quark production cross section at Fermilab was calculated by
Stange et.al \cite{willen}. In the framework of the SM, they found that the
 correction is less than $2.4 \%$ which is much less than the uncertainty in the cross
section. Also, they found that in the general two Higgs doublet model (2HDM), corrections can be
significantly enhanced, and can be as large as $20 \%$. 
On the other hand, for the minimal supersymmetric standard model (MSSM) 
the corrections of the Higgs sector never exceed $20 \%$. Finally, 
in connection to the cross section of the top quark production, the genuine supersymmetric electroweak corrections to $\bar{t}tg$ vertex  
were calculated in \cite{yang1, yang2}. 
The result of this calculation is  that for $\tan \beta >
1$ and the bottom-squark mass $<150$ GeV the correction can exceed $20 \%$.
Reference \cite{li} evaluated the supersymmetric QCD corrections to single
top quark production  and found that the combined effects of
supersymmetry and SM might reach the $10 \%$ for small $\tan \beta$.

 It has been pointed out that
$\Delta \kappa$ would be more easily probed at the LHC than the
Tevatron. 
Roughly speaking, the sensitivity of LHC can be a factor of three better
getting close to values
of $\Delta \kappa$ as small as 0.03 \cite{rizzo,silver}.
Non-QCD radiative corrections on the $t \bar t$
production at the LHC amounts to  $2.5 \%$  including the SM with a Higgs boson
mass $\sim 100$ GeV. For the general 2HDM the 
contribution of radiative corrections is less than $4 \%$, for
SUSY electroweak one-loop corrections less than $10 \%$ and if only SUSY QCD
one-loop corrections are considered, about  $4 \%$ \cite{hollik}. 

Since the anomalous chromomagnetic dipole moment of the top quark appears in
the top quark cross section, it is possible, due to uncertainties,  
to estimate the constraints that it would impose on the $\Delta \kappa$.
For the LHC, the
anomalous coupling is constrained to lie in the range $-0.09 \leq \Delta
\kappa \leq 0.1$ \cite{rizzo}. Similar range is obtained for the future NLC.  The influence of an anomalous $\Delta \kappa$ on the cross section
and associated gluon jet energy for $t \bar t g$ has 
been also analyzed. Events produced at 
$500$ GeV in $e^+ e^-$ colliders, with a cut on the gluon energy
of $500$ GeV and integrated luminosity of $30 fb^{-1}$,  lead to a
bound of $-2.1 \leq \Delta \kappa \leq 0.6$. Finally in
this context, Rizzo \cite{rizzo} showed that the top quark $p_t$ and $M_{tt}$ distributions
are the most sensitive observables to non-zero values of $\Delta \kappa$.

From the experimental information it is possible to get a
limit on the $\Delta \kappa$ from Tevatron. 
Following the reference by F. del Aguila \cite{exp} and  assuming that the only non-zero coupling is precisely the chromomagnetic dipole moment of the top quark, we find from the collected data that the allowed region is $\vert  \Delta \kappa \vert \leq 0.45$.

We call to the attention that although the vertex  $ttg$
is involved in the top quark pair production, the anomalous factor would be
measured at transferred momentum different from zero 
which, of course means an off-shell gluon.
Anomalous couplings of
the top quark to on-shell gluons would modify the rate for $B \to X_s \gamma$
\cite{hewet,us}. The presence of the magnetic dipole moment would affect the
Wilson coefficients which  mediate $b \to s $ transitions by the coefficients $C_{7,8}$ of the one-loop matching conditions
for the magnetic and chromomagnetic  dipole operators $O_{7,8}$. 
Comparing the
calculated branching fraction which involves the anomalous coupling, 
to the CLEO
measurements it is possible to get an allowed region for this anomalous coupling
\cite{hewet,us}. Using the recent data from CLEO collaboration for the
branching fraction of the process $B(b \to s
\gamma)=(3.21 \pm 0.43 \pm 0.27) \times 10^{-4}$ \cite{cleo}, we update the previous
analysis done in reference \cite{us} and get a new allowed region for the
anomalous chromomagnetic dipole moment of the top quark to be $-0.03 \leq \Delta
\kappa \leq 0.01$.

Our objective in this paper is to evaluate the contribution 
at the one loop-level to the anomalous
chromomagnetic dipole moment of the top quark in different scenarios with the
gluon boson on-shell. 
We study the region of allowed parameter space in different frameworks. 
We derive our bounds on  $\Delta \kappa$ 
from the analysis performed for the $b \to s \gamma$ process \cite{us}.

Beginning with the SM, the typical QCD correction through gluon exchange
implies two different Feynman diagrams: the first one is equivalent to the
QED contribution where the external gluon is coupled to the fermion line in
the loop and in the second one the external gluon is coupled to internal
gluons due to the non-abelian character of the $SU(3)$ color group. 
After the explicit calculation of the loops, we find that the second diagram
does not contribute and the final result is,
\begin{equation}
\Delta \kappa=-\frac 16 \frac{\alpha_s(m_t)}{\pi} 
\end{equation} 
We note that its
natural size is of the order of $\alpha_s/ \pi$ similar to the QED anomalous
coupling, but now 
in combination with a factor $-1/6$ coming from the color structure in the
diagram i.e.  $T^a T^b T^a=-T^b/6$ with $T^a$ 
being the generators of $SU(3)_C$. 

\begin{figure}[htbp] 
\begin{center}
\includegraphics[angle=0,width=7cm]{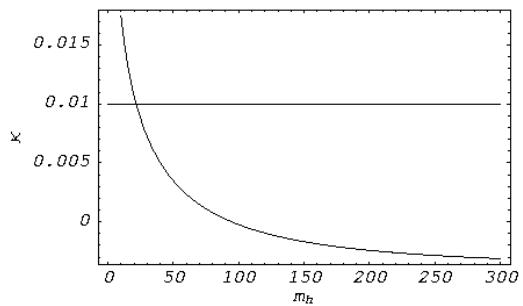}
\end{center}
\caption{Standard Model contribution to the anomalous chromomagnetic dipole
moment of the top quark versus the Higgs boson mass.}
\end{figure}

The other possible contribution in the framework of the SM comes
from electroweak interactions. The relevant contributions occur when neutral
Higgs boson and the would-be Goldstone boson of $Z$ are involved in the loop.
This contribution reads 
\begin{equation}
  \Delta \kappa= - \frac{\sqrt{2} G_F m_t^2}{8
\pi^2} [H_1(m_h)+H_2(m_Z)] ,  
\end{equation}
where
\begin{eqnarray}
H_1(m)&=& \int_0^1 dx \frac{x-x^3}{x^2-(2-m^2/m_t^2)x+1}  ,\nonumber \\
H_2(m)&=& \int_0^1 dx \frac{-x+2x^2-x^3}{x^2-(2-m^2/m_t^2)x+1}. \nonumber
\end{eqnarray}
This expression agree with similar one presented in reference \cite{peccei}. The SM contribution is showed in figure 1 where we have added the QCD
contribution (2). 
It is worth noting that the behaviour of the curve for a large Higgs
boson mass indicates decoupling and 
that the values of $\Delta \kappa$  lie within 
the allowed region for $\Delta \kappa$ coming from $b \to s \gamma$.

The contributions within a general 2HDM will be different
from the SM contributions because of the presence of the virtual five
physical Higgs bosons which appear in any two Higgs doublet model after
spontaneous symmetry breaking: $H^0$, $A^0$, $h^0$, $H^\pm$ \cite{hunter}.
Therefore, 2HDM predictions depend on their massses and
on the two mixing angles $\alpha$ and $\beta$. For small $\beta$, the charged
Higgs boson contribution is suppressed due to 
its large mass 
\cite{willen,rizzo} and the small bottom quark mass.

The expression  for the contribution of the neutral Higgs bosons is given by 
\begin{eqnarray} 
\Delta \kappa&=&\frac{\sqrt{2} G_F}{8 \pi^2} 
[\lambda_{H^0 tt}^2 H_1(M_H^2)+\lambda_{h^0 tt}^2 H_1(M_h^2)  \nonumber \\
&+&\lambda_{A^0tt}^2 H_2(M_A^2)+\lambda_{G^0 tt}^2 H_2(M_Z^2)]
\end{eqnarray}
where $\lambda_{itt}$ are the Yukawa couplings in the so-called models 
of type I,
II and III \cite{hunter}. Table I shows the couplings in the usual convention.
\begin{table}[htbp]
\begin{tabular}{l c c c }\\  
\hline
$\lambda_{itt}$ & model type I & model type II & model type III \\ \hline
$\lambda_{H^0 tt}$ & $\frac{m_t \sin \alpha}{\sin \beta}$ & $\frac{m_t \cos
\alpha}{\cos \beta}$ & $(1+\frac{\eta^U_{tt}}{\sqrt{2}})m_t \sin \alpha$\\ 
$\lambda_{h^0 tt}$ & $\frac{m_t \cos \alpha}{\sin \beta}$ & $\frac{m_t \sin
\alpha}{\cos \beta}$ & $(1+\frac{\eta^U_{tt}}{\sqrt{2}})m_t \cos \alpha$\\ 
$\lambda_{A^0tt}$ & $\cot \beta m_t$ & $\tan \beta m_t$ & $\frac{\eta^U_{tt}
m_t}{\sqrt{2}}$ \\ 
$\lambda_{G^0tt}$ & $m_t$ & $m_t$ & $m_t$ \\
\hline\\
\end{tabular}
\caption{Couplings of the Higgs eigenstates with the top quark which are
relevant in the calculation of the anomalous chromomagnetic dipole moment of
the top quark in 2HDM in its different versions. We
omit the factor $g/2 m_W$ and in the model type III we use the Sher-Cheng
approach for the flavour changing couplings, hence 
 the parameters $\eta_{ij}\sim 
1$ for the numerical analysis\cite{sher}.}
\end{table}

\begin{figure}[htbp]
\begin{center}
\includegraphics[angle=0,width=7cm]{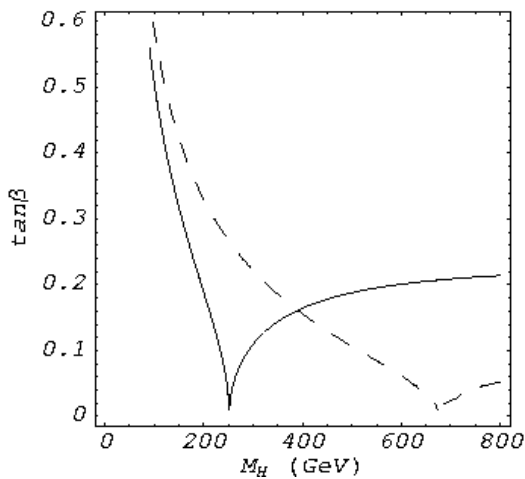}
\end{center}
\caption{Contour plot for the contribution of 2HDM to the anomalous chromomagnetic
dipole moment of the top quark  in the plane $\tan \beta-m_H$ for $m_A=200$ GeV (solid line) and $m_A=400$ GeV (dashed line)
using the bound from the $b \to s \gamma$
process. The allowed region is above the curve.} 
\end{figure}
The Yukawa couplings of a given fermion to the Higgs scalars are proportional
to the mass of the fermion and they are therefore naturally enhanced in this
case. In the model type III appears flavour changing neutral couplings at tree level which can be parametrized in the Sher-Cheng approach where a natural value for 
the flavour changing couplings from different families should be of the order of the 
geometric average of their Yukawa couplings, $h_{ij}=g \eta_{ij}\sqrt{m_i m_j}/(2 m_W)$ with $\eta_{ij}$ of the order of one \cite{sher}.

In order to show the behaviour of the contribution of the 2HDM to the anomalous chromomagnetic dipole moment of the top quark, we evaluate explicitly the contribution for couplings type II. We show in figure 2 the allowed region (above the curve)  for
the plane $\tan \beta$ vs $m_H$ using equation (4) and assuming that $\Delta \kappa$ is   $-0.03 \leq \Delta \kappa \leq  0.01$  from $b \to s \gamma$\cite{us}.  We fix the following parameters: $m_H=m_h$,  $m_A=200(400)$ GeV solid line (dashed line). The solid line for the scalar Higgs mass smaller (bigger) than $240$ GeV corresponds to the cut between equation (4) and the upper(lower) limit from $b \to s \gamma$. In figure 3 as in figure 2, we display the allowed region for the plane $\tan \beta$ vs $m_A$ with  $m_H=m_h=90(200)$ GeV for the solid (dashed) line. In this case we only find cuts with the upper limit for $\Delta \kappa$ from $b \to s \gamma$.  

\begin{figure}[ht]
\begin{center}
\includegraphics[angle=0,width=7cm]{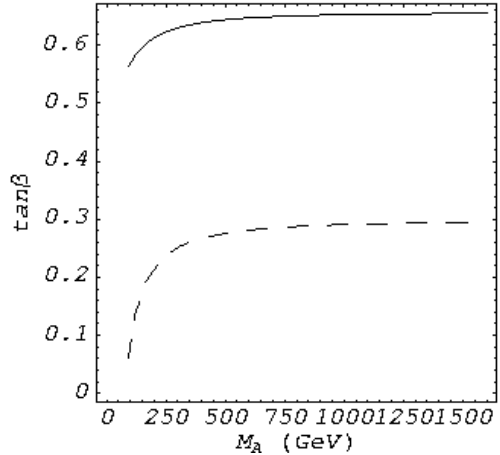}
\end{center}
\caption{Contour plot for the plane  $\tan \beta-m_A$ with the common scalar mass $m_h=m_H=90(200)$ GeV for the solid line (dashed line). The allowed region is above the curve. }
\end{figure}

Our last step is to calculate the anomalous chromomagnetic dipole moment
of the top quark in the framework of the MSSM. 
We only consider the SUSY QCD contribution which is generated from the
exchange of gluinos and squarks. 
The mixing in the squark sector in this model 
is particularly relevant in the case of
the squark top because the top quark mass is involved. In this case is
possible to neglect the family mixing and consider only the mixing between the
superpartners for the left- and right handed top quark. The mass eigenstates are given by
\begin{equation}
\tilde{t}_a=\sum_{b=ij} R_{ab} \tilde{q_b}
\end{equation}
with $R_{ab}$  the rotation matrix which diagonalizes the stop mass matrix

\begin{widetext}
\begin{equation}
M_{\tilde t}=\left( \begin{array}{cc}
\tilde M_Q +m_Z^2 \cos 2 \beta (\frac 12 - \frac 23 \sin^2 \theta_w)+m_t^2 &  (\mu \cot \beta +A_t \tilde M)m_t \\
(\mu \cot \beta +A_t \tilde M)m_t & \tilde M_U +m_Z^2 \cos 2 \beta (\frac 23
\sin^2 \theta_w)+m_t^2 
\end{array} \right)
\end{equation}
where $\tilde M_{Q,U}$ are the soft SUSY breaking terms, $\mu$ is the
coefficient  of the bilinear Higgs term and $A_t$ is the trilinear soft
SUSY breaking parameter. In this case, the mixing angle and the mass
eigenstates are
\begin{eqnarray}
\tan 2 \theta_t &=&\frac{2(\mu \cot \beta +A_t \tilde M)m_t}{\Delta \tilde
m_t} \\    
\tilde m_{t_{1,2}}&=&\frac 12 [\tilde m_{t_L}^2+\tilde m_{t_R}^2 \pm
\sqrt{(\Delta \tilde m_t)^2+4(\mu \cot \beta +A_t \tilde M)^2 m_t^2}]
\end{eqnarray}
where $\Delta \tilde m_t=\tilde m_{t_L}^2-\tilde m_{t_R}^2$ and $\tilde
m_{t_{L,(R)}}^2$ are the diagonal entries in the matrix (6).

In this framework the SUSY QCD contribution arises from the exchange of  
$\tilde t_{1,2}$ and gluinos. The contribution for the virtual $\tilde t_1$ 
turns out to be
\begin{eqnarray}  
\Delta \kappa &=& \frac{4 \alpha_s}{3 \pi} [\cos^2
\theta_t H_3(\hat {m_g},\hat {\tilde m_t}) -\sin^2 \theta_t H_4(\hat {m_g},\hat
{\tilde m_t})+\hat {m_g} \sin \theta_t \cos \theta_t H_5(\hat {m_g},\hat
{\tilde m_t})] \nonumber \\ 
&+& {\frac{3 \alpha_s}{2 \pi}} [\cos^2 \theta_t H_3(\hat {\tilde m_t},\hat
{m_g})+ \sin^2 \theta_t H_4(\hat {\tilde m_t},\hat {m_g})-\hat {m_g} \sin
\theta_t \cos \theta_t H_5(\hat {\tilde m_t},\hat {m_g})] 
\end{eqnarray}
\end{widetext}

\begin{figure}[h]
\begin{center}
\includegraphics[angle=0,width=7cm]{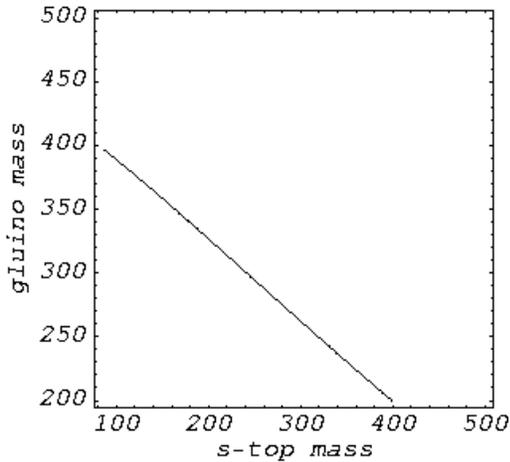}
\end{center}
\caption{Contour plot for the supersymmetric QCD contribution to the anomalous
chromomagnetic dipole moment of the top quark using the bound from the $b
\to s \gamma$ process. The allowed region is above the curve.} 
\end{figure}

with the functions
\begin{eqnarray}
H_3(m_1,m_2) &=& \int_0^1 dx \frac{x^2-3x^3/2}{x^2-(1-m_1^2+m_2^2)+m_2^2}
\nonumber \\ 
H_4(m_1,m_2) &=& \int_0^1 dx \frac{x^3/2}{x^2-(1-m_1^2+m_2^2)+m_2^2}
\nonumber \\
 H_5(m_1,m_2) &=& \int_0^1 dx \frac{x^2-x}{x^2-(1-m_1^2+m_2^2)+m_2^2} 
\end{eqnarray}
where $\hat m=m/m_t$.  For the squark $\tilde t_2$ the expression is
obtained by changing $\sin \theta_t$ to $\cos \theta_t$ and vice-versa.

Similar to figure 2 and 3, 
the plot in figure 4 refers to the allowed region of the parameter space  
obtained from constraints on $\Delta \kappa$ from $b \to s \gamma$.
Here  the plane is of the stop mass vs gluino mass with the SUSY parameters
fixed to $\theta_t=0.98$ and
$\tilde m_{t_1}=\tilde m_{t_2}$. The mixing angle is chosen such that the coupling of the squark top with the $Z^0$ boson vanishes and therefore we use the lower limits for the squark mass obtained from the LEP searching \cite{pdg}. The allowed region is above the curve and the solid line corresponds to the cut of  the upper limit of the bound of $\Delta \kappa$ coming from $b \to s \gamma$ and $\Delta \kappa$ from equation (9). 

In conclusion, the anomalous chromomagnetic dipole moment of the top quark 
receives the less stringent 
bound $\vert \Delta \kappa\vert \leq 0.45$ from the Tevatron experiments under the assumption that it is the only non-zero anomalous coupling. On the other hand, we found a more stringent bound from the transition $b \to s \gamma$, recently measured with improved precision by CLEO collaboration \cite{cleo}, and the bound is $-0.03 \leq \Delta \kappa \leq  0.01$. 
We have also calculated the chromomagnetic dipole moment of the top
quark in the framework of the SM, 2HDM and SUSY-QCD. We found that in SM the
anomalous coupling is up to the order of $\pm 10^{-2}$ and it is tending to $-4 \times 10^{-3}$  in the decoupling limit for 
a large Higgs mass. Furthermore, the anomalous coupling is equal to $-7 \times 10^{-4}$ around a Higgs mass of $113$ GeV which is the lower experimental limit from LEP2 \cite{pdg}.  If we keep in mind 
the sensitivity of future and 
current experiments to $\Delta \kappa$ $(\sim 0.03)$ \cite{rizzo} and its present
status, it is clear that 
an experimental measurement 
of this anomalous coupling  
could be a harbinger of physics beyond the SM.  

In the
2HDM the anomalous coupling of the top quark can reach $\sim 10^{-1}$ values, 
which is also possible for the
supersymmetric QCD corrections. We have compared the values of $\Delta \kappa$ obtained in  different scenarios with the bound from the
chromomagnetic dipole moment of the top quark resulting from $b \to s \gamma$
process. We have done this,  because  our considerations apply for on-shell gluons.
Therefore, we have presented a parameter analysis for different frameworks which allows an anomalous coupling of the same order that the bound obtained from the $b \to s \gamma$.  The allowed regions for parameters of the 2HDM type II are completely consistent with the unexplored experimental regions for $\tan \beta<1$ coming from analysis using data from LEP \cite{pdg}. Also, the allowed region for the plane $\tan \beta-$gluino mass agree with the experimental bounds.

We acknowledge to M. Nowakowski for the careful reading of the manuscript.
This work was supported by COLCIENCIAS, DIB and DINAIN.

\end{document}